\documentclass[journal]{IEEEtran}
\usepackage{stmaryrd}
\usepackage{amsmath}
\usepackage{amsthm}
\usepackage{xfrac}
\usepackage{amssymb}
\usepackage{mathabx}
\usepackage{psfrag}
\usepackage{graphicx}
\usepackage{epstopdf}
\usepackage{todonotes}
\usepackage{cite}
\usepackage{algorithmic}
\usepackage{algorithm}
\usepackage{svg}
\usepackage{mathtools}
\usepackage{hyperref}
\usepackage{hyphenat}
\usepackage{booktabs}
\usepackage{accents}
\usepackage{bbm}
\usepackage{balance}
\newlength{\dhatheight}

\usepackage{float}
\usepackage{multirow}

\usepackage{xparse,mathtools}

\newtheorem{corollary}{Corollary}
\newtheorem{lemma}{Lemma}

\begin{document}
%

\title{Maximally Concentrated Sequences after Half-sample Shifts}
\author{Karim A. Said, A. A. (Louis) Beex,  and Lingjia Liu
\thanks{K. Said, A. A. Beex and L. Liu are with Wireless@Virginia Tech, the Bradley Department of ECE at Virginia Tech, Blacksburg, VA. E. }
}
%




\maketitle
\begin{abstract}
It is well known that index (discrete-time)-limited sampled sequences leak outside the support set when a band-limiting operation is applied. Similarly, a fractional shift causes an index-limited sequence to be infinite in extent due to the inherent band-limiting. Index-limited versions of discrete prolate spheroidal sequences (DPSS) are known to experience minimum leakage after band-limiting. In this work, we consider the effect of a half-sample shift and provide upper bounds on the resulting leakage energy for arbitrary sequences. Furthermore, we find an orthonormal basis derived from DPSS with members ordered according to energy concentration \textit{after half sample shifts}; the primary (first) member being the global optimum.
\end{abstract}

\begin{IEEEkeywords}
Fractional shift, band-limit, index-limit
\end{IEEEkeywords}

\section{Introduction}
The shifting operation is intrinsic to physical systems: shifts in time such as delay in modeling of control systems \cite{1102802}, and delay due to wave propagation in wireless communication systems \cite{proakis2008digital}. Similarly, shifts can be in the frequency domain such as Doppler shifts \cite{proakis2017channel}, and space such as in sensor array processing \cite{10144727}. For discrete time systems, shifts that are integer multiples of the sample period are distortion-less operations; however, it is not the case for fractional shifts. A fractionally shifted version of a sequence can be thought of as a sampling of its shifted continuous version which is obtained by sinc function interpolation, i.e., band-limiting \cite{482137}. Band-limiting an index-limited sequence will result in leakage beyond the confines of the limited index support. Index limited discrete prolate spheroidal sequences (DPSS) are known to experience the least leakage when a band-limiting operation is applied; a property of DPSS known as optimal concentration. Index-limited DPSSs are eigenvectors of the composite operation of of band-limiting followed by index-limiting   \cite{slepian1978prolate}, which has given rise to numerous properties \cite{9647920} that are useful in many important signal processing applications\cite{1495893,9713839,8644309}.

A fractional shift can be thought of as a three step process: 1) Band-limited interpolation 2) Shift and 3) Sample. For any interpolation bandwidth, an index-limited DPSS exists that is optimally concentrated after this three step operation for integer sample shifts. However, this is \textit{not} true for fractional shifts. There are many examples of index-limited sequences that have favorable properties, such as sparsity, which are lost when they go "off-grid". This degradation is attributed to what is known as basis mismatch \cite{5710590}. One such example is harmonic complex sinusoidal sequences used in OFDM (Orthogonal Frequency Division Multiplexing) signals which are compromised in the presence of fractional shift channel effects \cite{yilmaz2022control}.  One motivation for this work was to find a set of orthogonal waveform vectors having a discrete index  support $[-N/2,...,N/2]$ that when shifted in time by $a \in \mathbb{R}^{+} \text{(causal shift)}$, the energy of the shifted versions is concentrated over the index set $[\lfloor-N/2\rfloor ,...,\lceil N/2+a\rceil ]$. It can be argued that the extreme case fractional shift is the case of $a=0.5$ \cite{482137} and hence will be the focus of this work. In pursuit of the aforementioned goal, the following contributions have been made:
\begin{itemize}
\item Theorem \ref{theorem1}: Upper bound on energy of leakage resulting from half-sample shifts for arbitrary sequences and a special case of equality;
\item Theorem \ref{theorem2}: Concentration of  arbitrary sequences after undergoing half-sample shifts;
\item Corollary \ref{corr_1}: Sequence with optimal concentration following half-sample shift;
\end{itemize}
in addition to supporting Lemmas 2, 3 and 4.

\subsection{Fractional delay operator}

We start by defining in \eqref{frac_shft} the fractional shift operator denoted  $\mathcal{B}_W^{\tau}$ which operates on a sequence $r[n]$ to generate an output sequence limited in frequency to half-bandwidth $W$ \text{Hz} (and scales by $\frac{1}{W}$) and shifted by $\tau \in \mathbb{R}$,

\begin{equation}\label{frac_shft}
\left\lbrace\mathcal{B}_W^{\tau} r\rbrace\right[n] \triangleq \sum_{q=-\infty}^{\infty} r[q]\text{sinc}\left(2W(n-\tau-q)\right)\\
\end{equation}
where $\text{sinc}(x)=\frac{\sin(\pi x)}{\pi x}$, and $0$ shift is denoted by $\mathcal{B}_W$.

Now we provide the following lemma.
\begin{lemma} \label{lemma1_statement}
Given a sequence $r[n]$,
\begin{equation} \label{lemma1_eq}
\lbrace\mathcal{B}_W^{0.5}r\rbrace[n]=\lbrace\mathcal{B}_{\frac{W}{2}}^{1}r\uparrow^2\rbrace[2n]
\end{equation}
where $\lbrace r\uparrow^2\rbrace[n] \triangleq r[\frac{n}{2}]  \text{ for even } n, 0 \text{ otherwise}$.

\end{lemma}
\textit{Proof}. 

Starting with the RHS of \eqref{lemma1_eq} and expanding using \eqref{frac_shft} 
\begin{equation}
\begin{split}
\lbrace\mathcal{B}_{\frac{W}{2}}^{1}\lbrace r\uparrow^2\rbrace\rbrace[2n]
&= \sum_{s=-\infty}^{\infty} \lbrace r\uparrow^2\rbrace[2s]\text{sinc}\left(2W\left(n-s-\frac{1}{2}\right)\right)\\
&=\sum_{s=-\infty}^{\infty} r[s]\text{sinc}\left(2W\left(n-s-\frac{1}{2}\right)\right)\\
&= \lbrace\mathcal{B}_W^{\frac{1}{2}}r\rbrace[n]\\
\end{split}  
\end{equation}
The significance of Lemma \ref{lemma1_statement} is that it places the $\tau=0.5$  fractional shift operation within the framework of discrete sequences with integer shifts, i.e., a fractional shift is replaced by an integer shift for an equivalent sequence. 
The following definition concerns measuring the leakage energy of an infinite sequence that results from a fractional shift of a finite length sequence.

\newtheorem{definition}{Definition}
\begin{definition}[Tail Energy]\label{def1}
  $\bar{E}_{-l,m}$ for $l,m\in \mathbb{Z}^{+}$:
 \begin{equation}\label{dbl_side_tail_E}
\bar{E}_{-l,m}(r)=\bar{E}_{-l,m}(r[n]) \triangleq \sum_{n=-\infty}^{-l-1}|r^2[n]|+\sum_{n=m+1}^{\infty}|r^2[n]|
\end{equation}
\end{definition}
\begin{lemma}\label{lemma2_statement}
For any complex sequence $r[n]$,
\begin{equation}\label{tail_E_inequality}
\bar{E}_{-l,m}(\mathcal{B}_W^{0.5}r)\leq \bar{E}_{-2l,2m}(\mathcal{B}_{0.5W}\lbrace r\uparrow^2\rbrace)-A
\end{equation}
where $A=0$ for infinite length sequences. For
finite length sequences $r[n]=0, n\notin -N/2,..,N/2, \quad N \text{ even}$, 
\begin{equation}\label{A_finite_seq}
A=\sum_{i=1}^{2}\sum_{\substack{q=-\frac{N}{2}}}^{\frac{N}{2}}  r[q]^2\text{sinc}^2\left(2W(-l-\frac{i}{2}-q)\right) 
\end{equation}
\end{lemma}
\textit{Proof}. 

Applying \eqref{lemma1_eq} and expanding in line $2$ and applying \eqref{dbl_side_tail_E}
\begin{equation}\label{lemma2_proof}
\begin{split}
&\bar{E}_{-l,m}(\mathcal{B}_W^{\frac{1}{2}}r)\\
&\overset{1}{=}\sum_{{\substack{n=-\infty, \\ even}}}^{-2l-2}\lbrace\mathcal{B}_{\frac{W}{2}}^{1}\lbrace r\uparrow^2\rbrace\rbrace^2[n]+\sum_{{\substack{n=2m+2, \\ even}}}^{-\infty}\lbrace\mathcal{B}_{\frac{W}{2}}^{1}\lbrace r\uparrow^2\rbrace\rbrace^2[n]\\
&\overset{2}{=}\sum_{{\substack{n=-\infty, \\ odd}}}^{-2l-3}\lbrace\mathcal{B}_{\frac{W}{2}}\lbrace r\uparrow^2\rbrace\rbrace^2[n]+\sum_{{\substack{n=2m+1, \\ odd}}}^{-\infty}\lbrace\mathcal{B}_{\frac{W}{2}}\lbrace r\uparrow^2\rbrace\rbrace^2[n]\\
&\overset{3}{\leq}   \sum_{-\infty}^{-2l-1}\lbrace\mathcal{B}_{\frac{W}{2}}\lbrace r\uparrow^2\rbrace\rbrace^2[n]  +\sum_{2m+1}^{-\infty}\lbrace\mathcal{B}_{\frac{W}{2}}\lbrace r\uparrow^2\rbrace\rbrace^2[n]
- A\\
\end{split}
\end{equation}
where $A=\sum_{i=1}^{2}\lbrace\mathcal{B}_{\frac{W}{2}}\lbrace r\uparrow^2\rbrace\rbrace^2[-2l-i]$. 
For finite length sequences $r[n]=0, n\notin -N/2,..,N/2$, $A$ involves a finite number of computations and by substituting using \eqref{frac_shft} it can be evaluated in \eqref{A_finite_seq} to tighten the bound given by \eqref{tail_E_inequality}.
\begin{lemma}\label{lemma3_statement}
For a finite length complex sequence $r[n]=0, n\notin -N/2,..,N/2 \quad N \text{ even}$, 
\begin{equation}\label{tail_E_inequality_lemma3}
\begin{split}
&\bar{E}_{-l,m}(\mathcal{B}_{\frac{1}{2}}^{\frac{1}{2}}r)= \bar{E}_{-2l-1,2m-1}(\mathcal{B}_{\frac{1}{4}}\lbrace r\uparrow^2\rbrace) \\
&\quad \forall l\geq \frac{N}{2}, m>\frac{N}{2}\\
\end{split}
\end{equation}
\end{lemma}
\textit{Proof}. 

Starting from line $2$ in \eqref{lemma2_proof} ,  setting $W=\frac{1}{2}$ and recognizing that the sequence $\lbrace\mathcal{B}_{\frac{1}{4}}\lbrace r\uparrow^2\rbrace\rbrace[n]=0 $ $\text{  for  }   \lbrace n\in\mathbb{R}-  [-(N+2),..,(N+2)], n \text{ even}\rbrace$ 
\begin{equation}\label{lemma3_proof_2}
\begin{split}
&\bar{E}_{-l,m}(\lbrace\mathcal{B}_{\frac{1}{2}}^{\frac{1}{2}}r\rbrace[n])\\
&\overset{1}{=}\sum_{{\substack{n=-\infty, \\ odd}}}^{-2l-3}\lbrace\mathcal{B}_{\frac{1}{4}}\lbrace r\uparrow^2\rbrace\rbrace^2[n]+\sum_{{\substack{n=2m+1, \\ odd}}}^{-\infty}\lbrace\mathcal{B}_{\frac{1}{4}}\lbrace r\uparrow^2\rbrace\rbrace^2[n]\\
&\overset{2}{=}\sum_{{\substack{n=-\infty}}}^{-2l-2}\lbrace\mathcal{B}_{\frac{1}{4}}\lbrace r\uparrow^2\rbrace\rbrace^2[n]+\sum_{{\substack{n=2m}}}^{-\infty}\lbrace\mathcal{B}_{\frac{1}{4}}\lbrace r\uparrow^2\rbrace\rbrace^2[n]\\
&\quad \forall l\geq \frac{N}{2}, m>\frac{N}{2}\\
\end{split}
\end{equation}
We list the following known properties about DPSS that will be useful in the following Lemma.

For $N \text{  even}$, a DPSS $s_l[n;2N+3,0.25]$ and its corresponding eigenvalue $\lambda_l(2N+3,0.25)$ $l=0,..,2N+2$ satisfy the following properties [(24), (23) respectively in \cite{slepian1978prolate}]:
\begin{equation}\label{DPSS_flip}
s_l[n;2N+3,0.25]=(-1)^{n+1}s_{2N+2-l}[-n;2N+3,0.25] 
\end{equation} 
\begin{equation}\label{lambda_flip}
\lambda_l(2N+3,0.25)=1-\lambda_{2N+2-l}(2N+3,0.25)
\end{equation}
for $n=-(N+1),..,(N+1) \text{ for } l=0,..,2N+2, l\neq N+1$.
\begin{lemma}\label{lemma4_statement}
For $N \text{  even}$, the set of DPSS sequences $\sqrt{2}s_l[2n;2N+3,0.25], n=-N/2,..,N/2, l=0,..,N$  forms an orthonormal basis.
\end{lemma}
\textit{Proof}. 

By DPSS orthogonality:
\begin{equation}\label{DPSS_perp}
\begin{split}
&\sum_{-(N+1)}^{N+1} 
s_l[n]s_k[n]\\
&=\sum_{n=-\frac{N}{2}}^{\frac{N}{2}}s_l[2n]s_k[2n]+ \sum_{n=-\frac{N+2}{2}}^{\frac{N}{2}}s_l[2n+1]s_k[2n+1]=0\\
\end{split}
\end{equation}
Without loss of generality, we start with the case $l=2p \text{ (even)}, k=2q+1\text{ (odd)}$ and substitute in \eqref{DPSS_perp}
\begin{equation}\label{DPSS_perp2}
\begin{split}
&\sum_{-\frac{N}{2}}^{\frac{N}{2}}s_{2p}[2n]s_{2q+1}[2n]+\sum_{-\frac{N+2}{2}}^{N/2}s_{2p}[2n+1]s_{2q+1}[2n+1]=0\\
\end{split}
\end{equation}
Since $s_{2p}[2n]s_{2q+1}[2n]$ is an odd sequence $\sum_{-\frac{N}{2}}^{\frac{N}{2}}s_{2p}[2n]s_{2q+1}[2n]=0$ and therefore $\sum_{-\frac{N+2}{2}}^{N/2}s_{2p}[2n+1]s_{2q+1}[2n+1]=0$.

By \eqref{DPSS_flip}, \eqref{DPSS_perp} is simultaneously true for $l, k$ and $l, 2N+2-k$, $l, k< N+1$ therefore we can write:
\begin{equation}\label{ee_oo_case}
\begin{split}
&\sum_{-\frac{N}{2}}^{\frac{N}{2}}s_{l}[2n](s_{k}[2n]+(-1)^{2n+1+k}s_{k}[2n])\\
&+ \sum_{-(\frac{N+2}{2})}^{\frac{N}{2}}s_{l}[2n+1](s_{k}[2n+1]+(-1)^{2n+2+k}s_{k}[2n+1])\\
&\sum_{-\frac{N}{2}}^{\frac{N}{2}}s_{l}[2n](s_{k}[2n]-(-1)^ks_{k}[2n])\\
&+ \sum_{-(\frac{N+2}{2})}^{\frac{N}{2}}s_{l}[2n+1](s_{k}[2n+1]+(-1)^ks_{k}[2n+1])=0
\end{split}
\end{equation}
For $(l,k)=(2p,2q), p\neq q$, by substituting in \eqref{ee_oo_case}  we get $2\sum_{-(\frac{N+2}{2})}^{\frac{N}{2}}s_{2p}[2n+1]s_{2q}[2n+1]=0$, and consequently $\sum_{-(\frac{N+2}{2})}^{\frac{N}{2}}s_{2p}[2n]s_{2q}[2n]=0$. Similarly, for $(l,k)=(2p+1,2q+1), p\neq q$, $2\sum_{-\frac{N}{2}}^{\frac{N}{2}}s_{2p+1}[2n]s_{2q+1}[2n]=0$ and $\sum_{-\frac{N+2}{2}}^{\frac{N}{2}}s_{2p+1}[2n+1]s_{2q+1}[2n+1]=0$.
Finally, for $l=k<N+1$, $2\sum_{-\frac{N+2}{2}}^{\frac{N}{2}}s_{l}[2n+1]s_{l}[2n+1]=1 \text{ for even } l$ and $2\sum_{-\frac{N+2}{2}}^{\frac{N}{2}}s_{l}[2n]s_{l}[2n]=1 \text{ for odd } l$ .

\newtheorem{theorem}{Theorem}
\begin{theorem}[Tail energy of half-sample shifted sequences]\label{theorem1}
For a complex sequence $r[n]$ where $r[n] = 0 \text{ for } n \notin -N/2,..,N/2$, we have the following upper-bound on the tail energy:
\begin{equation}\label{tail_E_bound}
\bar{E}_{-N/2,N/2}(\mathcal{B}_W^{0.5}r)\leq \sum_{l=0}^{2N}\left|\frac{a_l}{W}\right|^2\lambda_l(1-\lambda_l)
\end{equation}
where $a_l=\sum_{n=-N/2}^{N/2}r[n]s_l^{(0.5W,2N+1)}[2n]$, $\lambda_l$ is the eigenvalue of the $l$-th member  $s_l^{(0.5W,2N+1)}$  from the DPSS set with parameters $0.5W,2N+1$.

For the special case of $W=0.5$, we have the following equality
\begin{equation}\label{tail_E_bound_equality}
\bar{E}_{-N/2,N/2+1}(\mathcal{B}_{0.5}^{0.5}r) = 8\sum_{l=0}^{N} \left|\bar{a}_l\right|^2\bar{\lambda}_l(1-\bar{\lambda}_l)
\end{equation}
 and where $\bar{a}_l=\sum_{n=-N/2}^{N/2}r[n]s_l^{(0.25,2N+3)}[2n]$, $\bar{\lambda}_l$ is the eigenvalue of the $l$-th member  $s_l^{(0.25,2N+3)}$  from the DPSS set with parameters $0.25,2N+3$.
\end{theorem}
\textit{Proof}. 

For a DPSS set with parameters: $2N+1,0.5W$, $s_l[n]$ is assumed to be normalized according to \eqref{normalization}
\begin{equation} \label{normalization}
\sum_{m=-N}^{N} s^2_l[m] =1
\end{equation}
By the property of  spectral concentration of DPSS \eqref{spec_conc} which can be found in \cite{slepian1978prolate}:
\begin{equation} \label{spec_conc}
\frac{   \int_{-0.5W}^{0.5W}\lvert \mathcal{S}_l(f)\rvert^2df  }{ \int_{-0.5}^{0.5}\lvert \mathcal{S}_l(f)\rvert^2df }=\frac{   \int_{-0.5W}^{0.5W}\lvert \mathcal{S}_l(f)\rvert^2df  }{ \sum_{m=-N}^{N} s^2_l[m]}=\lambda_l
\end{equation}
where $S_l(f)=\sum_{n=-N}^{N}s_l[n]e^{j2\pi f n}$ is the discrete Fourier transform and $\int_{-0.5}^{0.5}|S_l(f)|^2df=1$.
Writing the defining eigen-function equation of DPSS \cite{slepian1978prolate} in terms of our operator notation $\mathcal{B}$
\begin{equation} \label{Dpss_efunc_eq}
\begin{split}
\lambda_ls_l[n] &= W\mathcal{B}_{0.5W}s_l[n]
\end{split}
\end{equation}
Computing energy using the right hand expression of \eqref{Dpss_efunc_eq}:
\begin{equation}\label{full_energy}
\begin{split}
&\sum_{-\infty}^{\infty}\lbrace\mathcal{B}_{\frac{W}{2}}s_l\rbrace^2[n]\\
&\overset{1}{=}\sum_{-\infty}^{-N-1}\lbrace\mathcal{B}_{\frac{W}{2}}s_l\rbrace^2[n]+\sum_{N+1}^{\infty}\lbrace\mathcal{B}_{\frac{W}{2}}s_l\rbrace^2[n]+\sum_{-N}^{N}\lbrace\mathcal{B}_{\frac{W}{2}}s_l\rbrace^2[n]\\
&\overset{2}{=}\sum_{-\infty}^{-N-1}\lbrace\mathcal{B}_{\frac{W}{2}}s_l\rbrace^2[n]+\sum_{N+1}^{\infty}\lbrace\mathcal{B}_{\frac{W}{2}}s_l\rbrace^2[n]+\left(\frac{\lambda_l}{W}\right)^2\sum_{-N}^{N}s_l^2[n]\\
&\overset{3}{=}\sum_{-\infty}^{-N-1}\lbrace\mathcal{B}_{\frac{W}{2}}s_l\rbrace^2[n]+\sum_{N+1}^{\infty}\lbrace\mathcal{B}_{\frac{W}{2}}s_l\rbrace^2[n]+\left(\frac{\lambda_l}{W}\right)^2\\
\end{split}
\end{equation}
Using the frequency domain representation of the right-hand-side (RHS) of \eqref{Dpss_efunc_eq} and applying  \eqref{spec_conc} 
\begin{equation}\label{E_bandlimited}
W^2\sum_{-\infty}^{\infty}\lbrace\mathcal{B}_{0.5W}s_l\rbrace^2[n]=\int_{-0.5W}^{0.5W}\lvert \mathcal{S}_l(f)\rvert^2df = \lambda_l
\end{equation}
Substituting \eqref{E_bandlimited} into the left hand side of \eqref{full_energy} and rearranging terms:
\begin{equation}\label{double_sided_tail_E}
\begin{split}
\sum_{-\infty}^{-N-1}\lbrace\mathcal{B}_{0.5W}s_l\rbrace^2[n]+\sum_{N+1}^{\infty}\lbrace\mathcal{B}_{0.5W}s_l\rbrace^2[n]&=\frac{\lambda_l -\lambda_l^2}{W^2}\\
&=\bar{E}_{-N,N}(\mathcal{B}_{0.5W}s_l)
\end{split}
\end{equation}
The set of index-limited DPSS denoted $\mathcal{I}_{2N+1}s_l$ defined by \eqref{index_limited_DPSS} is a complete basis for the space of index-limited sequences over $-N,..,N$
\begin{equation}\label{index_limited_DPSS}
\mathcal{I}_{2N+1}s_l[n]\triangleq\begin{cases}
                        s_l[n] \quad n \in -N,..,N   \\
                        0 \text{ otherwise }
                    \end{cases}
\end{equation}
Thus, an arbitrary complex sequence $\tilde{r}[n]$,  $\tilde{r}[n] = 0 \text{ for } n \notin -N,..,N$, admits the representation \eqref{basis_expansion}
\begin{equation}\label{basis_expansion}
\tilde{r}[n]=\sum_{l=0}^{2N} \tilde{a}_l \mathcal{I}_{2N+1}s_l[n]
\end{equation}
where $\tilde{a}_l=\sum_{n=-N}^{N} s_l[n]\tilde{r}[n]$.
Finding the tail energy of the sequence $\mathcal{B}_{0.5W}\tilde{r}$ and using \eqref{double_sided_tail_E}
\begin{equation}\label{basis_expansion_tailE}
\begin{split}
\bar{E}_{-N,N}(\mathcal{B}_{0.5W}\tilde{r})&=\sum_{l=0}^{2N} \left|\frac{\tilde{a}_l}{W}\right|^2\bar{E}_{-N,N}(\mathcal{B}_{0.5W}s_l)\\
&=\sum_{l=0}^{2N} \left|\frac{\tilde{a}_l}{W}\right|^2\lambda_l(1-\lambda_l)
\end{split}
\end{equation}
Applying \eqref{tail_E_inequality} from Lemma \ref{lemma2_statement} to a sequence $r[n]$ where ${r}[n] = 0 \text{ for } n \notin -\frac{N}{2},..,\frac{N}{2}$, and substituting in \eqref{basis_expansion_tailE} by $\tilde{r}=r\uparrow^2$
\begin{equation}
\begin{split}
\bar{E}_{-\frac{N}{2},\frac{N}{2}}(\mathcal{B}_W^{0.5}r)&\leq \bar{E}_{-N,N}(\mathcal{B}_{0.5W}\lbrace r\uparrow^2\rbrace) \\
& = \sum_{l=0}^{2N} \left|\frac{a_l}{W}\right|^2\lambda_l(1-\lambda_l)
\end{split}
\end{equation}
where $a_l=\sum_{n=-N}^{N} s_l[n]\lbrace r\uparrow^2\rbrace[n]=\sum_{n=-N/2}^{N/2} s_l[2n] r[n]$,   $\lambda_l$ is the eigenvalue of the $l$-th member  $s_l^{(0.5W,2N+1)}$  from the DPSS set with parameters $0.5W,2N+1$.

For the special case of $W=0.5$, we make use of \eqref{lemma3_proof_2} from Lemma \ref{lemma3_statement} and then substitute using \eqref{basis_expansion_tailE}

\begin{equation}\label{tail_E_inequality_lemma3}
\begin{split}
\bar{E}_{-\frac{N}{2},\frac{N}{2}+1}(\mathcal{B}_{0.5}^{0.5}r)&= \bar{E}_{-(N+1),N+1}(\mathcal{B}_{0.25}\lbrace r\uparrow^2\rbrace) \\
& = 4\sum_{l=0}^{2N+2} \left|\bar{a}_l\right|^2\bar{\lambda}_l(1-\bar{\lambda}_l)
\end{split}
\end{equation}
where $\bar{a}_l=\sum_{n=-N/2}^{N/2} s_l[2n;0.25,2N+3] r[n]$, $\bar{\lambda}_l$ is the eigenvalue of the $l$-th member from the DPSS set with parameters $0.25,2N+3$, i.e., $s_l[n;0.25,2N+3]$.

By \eqref{DPSS_flip}, $|\bar{a}_l|=|\bar{a}_{2N+2-l}|$, in addition $s_{N+1}[2n;0.25,2N+3]=0$ \footnote{can be proven based on Lemma 4}  and therefore $\bar{a}_{N+1}=0$. Substituting in \eqref{tail_E_inequality_lemma3}
\begin{equation}\label{tail_E_inequality_lemma3_reduced}
\begin{split}
\bar{E}_{-\frac{N}{2},\frac{N}{2}+1}(\mathcal{B}_{0.5}^{0.5}r)
& = 8\sum_{l=0}^{N} \left|\bar{a}_l\right|^2\bar{\lambda}_l(1-\bar{\lambda}_l)
\end{split}
\end{equation}

\begin{theorem}[Concentration of Half-sample Shifted Sequences]\label{theorem2}
For an index limited complex sequence $r[n]$ where $r[n] = 0 \text{ for } n \notin -\frac{N}{2},..,\frac{N}{2}$, the concentration of $r[n]$ after undergoing a half sample shift, i.e.,  $\mathcal{B}_{0.5}^{0.5}r[n]$ is defined and evaluated as follows:

\begin{equation}\label{half_shift_conc}
\frac{E_{-\frac{N}{2},\frac{N}{2}+1}(\mathcal{B}_{0.5}^{0.5}r) }{E(\mathcal{B}_{0.5}^{0.5}r)} = \frac{1}{ 1+\frac{\sum_{l=0}^{N} \left|\bar{a}_l\right|^2\bar{\lambda}_l(1-\bar{\lambda}_l)}{\sum_{l=0}^{N}\bar{\lambda}_l^2|\bar{a}_l|^2-\frac{1}{8}}}
\end{equation}
where $\bar{a}_l=\sum_{n=-N/2}^{N/2}r[n]s_l^{(0.25,2N+3)}[2n]$, and $\bar{\lambda}_l$ is the eigenvalue of the $l$-th member  $s_l^{(0.25,2N+3)}$  from the DPSS set with parameters $0.25,2N+3$.
 \end{theorem}
\textit{Proof}. 

We expand the LHS of \eqref{half_shift_conc} as follows:

\begin{equation}\label{half_shift_conc_LHS}
\frac{E_{-\frac{N}{2},\frac{N}{2}+1}(\mathcal{B}_{0.5}^{0.5}r) }{E(\mathcal{B}_{0.5}^{0.5}r)}=\frac{E_{-\frac{N}{2},\frac{N}{2}+1}(\mathcal{B}_{0.5}^{0.5}r) }{E_{-\frac{N}{2},\frac{N}{2}+1}(\mathcal{B}_{0.5}^{0.5}r)+\bar{E}_{-\frac{N}{2},\frac{N}{2}+1}(\mathcal{B}_{0.5}^{0.5}r)}
\end{equation}
The second term in the denominator $\bar{E}_{-\frac{N}{2},\frac{N}{2}+1}(\mathcal{B}_{0.5}^{0.5}r)$ can be evaluated by  \eqref{tail_E_bound_equality} from Theorem 1. Now we need to evaluate the numerator term.
Following a similar approach to line-3 in \eqref{lemma2_proof}
\begin{equation}\label{mid_energy}
\begin{split}
&E_{-\frac{N}{2},\frac{N}{2}+1}(\lbrace\mathcal{B}_{0.5}^{0.5}r\rbrace[n])\\
&{=}\sum_{{\substack{-(N+1), \\ odd}}}^{N+1}\lbrace\mathcal{B}_{0.25}\lbrace r\uparrow^2\rbrace\rbrace^2[n]\\
&{=}\sum_{-(N+1)}^{N+1}\lbrace\mathcal{B}_{0.25}\lbrace r\uparrow^2\rbrace\rbrace^2[n]-\sum_{{\substack{-(N+1), \\ even}}}^{N+1}\lbrace\mathcal{B}_{0.25}\lbrace r\uparrow^2\rbrace\rbrace^2[n]\\
&{=}\sum_{-(N+1)}^{N+1}\lbrace\mathcal{B}_{0.25}\lbrace r\uparrow^2\rbrace\rbrace^2[n]-\sum_{-N/2}^{N/2}\lbrace\mathcal{B}_{0.5}\lbrace r\uparrow^2\rbrace\rbrace^2[2n]\\
&{=}\sum_{-(N+1)}^{N+1}\lbrace\mathcal{B}_{0.25}\lbrace r\uparrow^2\rbrace\rbrace^2[n]-\sum_{-N/2}^{N/2}r^2[n]\\
&{=}\sum_{-(N+1)}^{N+1}\lbrace\mathcal{B}_{0.25}\lbrace r\uparrow^2\rbrace\rbrace^2[n]-1\\
&{=}4\sum_{-(N+1)}^{N+1}\left\lbrace \frac{1}{2}\mathcal{B}_{0.25}\left\lbrace \sum_{l=0}^{2N+2}\bar{a}_ls_l[n;0.25,2N+3]\right\rbrace\right\rbrace^2[n]-1\\
&=8\sum_{l=0}^{N}(\bar{\lambda}_l\bar{a}_l)^2-1\\
\end{split}
\end{equation}

Substituting in the RHS of \eqref{half_shift_conc_LHS}, using \eqref{mid_energy} in the numerator and the left term of the denominator, and substituting using \eqref{tail_E_bound_equality} in the right term of the denominator of \eqref{half_shift_conc_LHS}
\begin{equation}\label{half_shift_conc_eval}
\begin{split}
&\frac{E_{-\frac{N}{2},\frac{N}{2}+1}(\mathcal{B}_{0.5}^{0.5}r) }{E(\mathcal{B}_{0.5}^{0.5}r)}\\
&=\frac{8\sum_{l=0}^{N}\bar{\lambda}_l^2|\bar{a}_l|^2-1}{ 8\sum_{l=0}^{N}\bar{\lambda}_l^2|\bar{a}_l|^2-1+8\sum_{l=0}^{N} \left|\bar{a}_l\right|^2\bar{\lambda}_l(1-\bar{\lambda}_l)}\\
&=\frac{1}{ 1+\frac{\sum_{l=0}^{N} \left|\bar{a}_l\right|^2\bar{\lambda}_l(1-\bar{\lambda}_l)}{\sum_{l=0}^{N}\bar{\lambda}_l^2|\bar{a}_l|^2-\frac{1}{8}}}\\
\end{split}
\end{equation}

\begin{corollary}\label{corr_1}
For complex sequences $r[n]$ with index support $[-N/2,N/2], \text{ } N\text{ even}$, the sequence having optimum concentration over support $[-N/2,N/2+1]$ after a half-sample fractional shift is $s_0[2n;2N+3,0.25]$, with concentration
\begin{equation}\label{half_shift_conc_eval}
\begin{split}
\frac{E_{-N/2,N/2+1}(\mathcal{B}_{0.5}^{0.5}r) }{E(\mathcal{B}_{0.5}^{0.5}r)}
=\frac{1}{ 1+\frac{ \bar{\lambda}_0(1-\bar{\lambda}_0)}{\bar{\lambda}_0^2-\frac{1}{8}}}\\
\end{split}
\end{equation}
\end{corollary}
\textit{Proof}. 

We rewrite \eqref{half_shift_conc} as follows:

\begin{equation}\label{half_shift_conc_matrix_form}
\frac{E_{-N/2,N/2+1}(\mathcal{B}_{0.5}^{0.5}r) }{E(\mathcal{B}_{0.5}^{0.5}r)} = \frac{1}{ 1+\frac{||(\boldsymbol{\lambda}\mathbf{J}\boldsymbol{\lambda})^{\frac{1}{2}}\mathbf{S}^T\mathbf{r}||_2^2}{||(\boldsymbol{\lambda}\boldsymbol{\lambda})^{\frac{1}{2}}\mathbf{S}^T\mathbf{r}||_2^2-\frac{1}{8}}}
\end{equation}
where $\lbrace\mathbf{S}\rbrace_{n,m}=s_c[2r;2N+3,0.25]$, $\lbrace\boldsymbol{\lambda}\rbrace_{n,m}=\lambda_{m+N/2}(2N+3,0.25)\delta(n-m)$, $\lbrace\mathbf{J}\rbrace_{n,m}=\delta(n+m)$, $\lbrace\mathbf{r}\rbrace_n=r[n]$ $\quad n,m= -N/2,..,N/2$.

To find the optimally concentrated sequence after a half sample shift, we attempt to solve the following optimization problem:

\begin{equation}\label{optimal_r}
\begin{split}
&\arg \max_{\mathbf{r}} \frac{1}{ 1+\frac{||(\boldsymbol{\lambda}\mathbf{J}\boldsymbol{\lambda})^{\frac{1}{2}}\mathbf{S}^T\mathbf{r}||_2^2}{||(\boldsymbol{\lambda}\boldsymbol{\lambda})^{\frac{1}{2}}\mathbf{S}^T\mathbf{r}||_2^2-\frac{1}{8}}}, \quad ||\mathbf{r}||_2=1\\
&\overset{1}{=} \arg \min_{\mathbf{r}}\frac{||(\boldsymbol{\lambda}\mathbf{J}\boldsymbol{\lambda})^{\frac{1}{2}}\mathbf{S}^T\mathbf{r}||_2^2}{||(\boldsymbol{\lambda}\boldsymbol{\lambda})^{\frac{1}{2}}\mathbf{S}^T\mathbf{r}||_2^2-\frac{1}{8}}\\
&\overset{2}{=}\arg \min_{\mathbf{r}}||(\boldsymbol{\lambda}\mathbf{J}\boldsymbol{\lambda})^{\frac{1}{2}}\mathbf{S}^T\mathbf{r}||_2^2=\arg \max_{\mathbf{r}}||(\boldsymbol{\lambda}\boldsymbol{\lambda})^{\frac{1}{2}}\mathbf{S}^T\mathbf{r}||_2^2\\
&\overset{3}{=}\mathbf{S}(0,:)
\end{split}
\end{equation}
By \eqref{lambda_flip}, $\boldsymbol{\lambda}\boldsymbol{I}\boldsymbol{\lambda}$ and $\boldsymbol{\lambda}$  are diagonal matrices with monotonically increasing and decreasing elements respectively. In addition, by the result of Lemma \ref{lemma4_statement},  $\mathbf{S}$ is an orthogonal matrix. As a result,  the maximizer of the numerator of line-1 simultaneously minimizes the denominator of line-1 as indicated by line-2.

\section{Conclusion}
This work addressed the effect of fractional shifts on index-limited sequences. Specifically, the focus was directed to shifts by half a sample which can be considered the worst case fractional shift.  Even though the resulting leakage is of infinite extent, using the DPSS framework the problem is reduced to a finite dimensional one enabling us to obtain bounds on the leakage energy that can be evaluated using a finite number of computations. Finally, the sequence of optimum concentration following a half sample shift was obtained, which was shown to be a sub-sampled DPSS sequence from the set of DPSSs of double the length of the energy concentration interval.

\IEEEpeerreviewmaketitle


%

\bibliographystyle{IEEEtran}
\bibliography{IEEEabrv,references.bib}

\end{document}